\documentclass{article}

\usepackage[T1]{fontenc}
\usepackage{graphicx}
\usepackage{tabularx}
\usepackage{booktabs}
\usepackage{tcolorbox}
\usepackage{listings}
\usepackage{color}
\usepackage[all]{nowidow}
\usepackage{amsmath,amssymb,amsfonts}
\usepackage{colortbl}
\usepackage{xspace}
\usepackage{mdframed}
\usepackage{authblk} 
\usepackage[absolute,overlay]{textpos}
\usepackage[hyphens]{url}
\usepackage[colorlinks=false, breaklinks]{hyperref}

\newcolumntype{Y}{>{\centering\arraybackslash}X}

\newcommand{\SISW}{SISW\xspace}

\definecolor{dkgreen}{rgb}{0,0.6,0}
\definecolor{gray}{rgb}{0.5,0.5,0.5}
\definecolor{mauve}{rgb}{0.58,0,0.82}

\lstdefinestyle{MR}{frame=tb,
	language=Java,
	aboveskip=3mm,
	belowskip=3mm,
	showstringspaces=false,
	columns=flexible,
	basicstyle={\scriptsize\ttfamily},
	numbers=left,
	xleftmargin=2em,
	numberstyle=\tiny\color{gray},
	keywordstyle=\color{blue},
	commentstyle=\color{dkgreen},
	stringstyle=\color{mauve},
	breaklines=true,
	breakatwhitespace=true,
	tabsize=2
}

\begin{document}

\begin{textblock*}{\textwidth}(4.5cm,3cm)
\footnotesize
This preprint has not undergone peer review (when applicable) or any post-submission improvements or corrections. The Version of Record of this contribution is published in Communications in Computer and Information Science (CCIS, volume 2178), and is available online at \url{https://doi.org/10.1007/978-3-031-70245-7_9}.
\end{textblock*}

\title{Towards Generating Executable Metamorphic Relations Using Large Language Models}

\author[1]{Seung Yeob Shin}
\author[1]{Fabrizio Pastore}
\author[1]{Domenico Bianculli}
\author[2]{Alexandra Baicoianu}

\affil[1]{SnT Centre, University of Luxembourg, Luxembourg
	{\{seungyeob.shin, fabrizio.pastore, domenico.bianculli\}@uni.lu}}
\affil[2]{Siemens Industry Software, Romania
	{alexandra.baicoianu@siemens.com}}

\date{} 

\maketitle

\begin{abstract}
Metamorphic testing (MT) has proven to be a successful solution to automating testing and addressing the oracle problem.
However, it entails manually deriving metamorphic relations (MRs) and converting them into an executable form; these steps are time-consuming and may prevent the adoption of MT.
In this paper, we propose an approach for automatically deriving executable MRs (EMRs) from requirements using large language models (LLMs).
Instead of merely asking the LLM to produce EMRs, our approach relies on a few-shot prompting strategy to instruct the LLM to perform activities in the MT process, by providing requirements and API specifications, as one would do with software engineers. 
To assess the feasibility of our approach, we conducted a questionnaire-based survey in collaboration with 
Siemens Industry Software,
a worldwide leader in providing industry software and services, focusing on four of their software applications. Additionally, we evaluated the accuracy of the generated EMRs for a Web application. The outcomes of our study are highly promising, as they demonstrate the capability of our approach to generate MRs and EMRs that are both comprehensible and pertinent for testing purposes.

\textbf{keywords:} metamorphic testing, large language model, LLM, executable metamorphic relations
\end{abstract}

 \section{Introduction}
\label{sec:intro}

In many sectors, software is typically verified with executable test cases that, at a high level, consist of a set of inputs provided to the software under test (SUT) and a set of test assertions verifying that the SUT outputs match the expected results. 
However, defining executable test cases is expensive because of the many intellectual activities involved. 
This cost limits the number of test cases that can be implemented. 
Moreover, software faults are often subtle and triggered by a narrow portion of the input domain.
Therefore, they are detected only after \hbox{exercising the SUT with a large set of test inputs.}

Although several tools for the automated generation of test inputs have been developed~\cite{AFL++,EvoSuite}, they can only identify crashes or regressions faults, because they lack the capability of determining what the expected output for the software should be. Further, manually specifying test assertions for thousands of automatically generated test inputs is practically infeasible. 
The impossibility to programmatically derive expected outputs is known as the \emph{oracle problem}~\cite{Barr:Oracles:15}.

Metamorphic testing (MT) has recently gained success as a solution to address the oracle problem in many contexts~\cite{Chen2018}.
MT relies on the concept of metamorphic relations (MRs), which describe the relationships between input transformations and expected output changes, serving as oracles in software testing.
Briefly, to perform MT on the SUT, engineers first need to derive MRs from the SUT's requirements and then convert the derived MRs into an executable form to determine the test outcome. 
These steps require substantial manual effort, which increases the MT cost and may prevent its adoption.

Many MT approaches have been developed to test various SUTs, including search engines~\cite{Zhou2016}, Web applications~\cite{Chaleshtari:MST}, and embedded systems~\cite{Ayerdi:2020}.
These methods, however, require significant manual efforts to define MRs, which limits their scalability, particularly in cases where SUT executions are lengthy and expensive.
A few recent studies~\cite{luu2023chatgpt,Zhang:MRs:2023} have investigated using Large Language Models (LLMs) to automate MR derivation directly from LLM knowledge bases.
However, they currently only work with SUTs known during LLM training and are not effectively applicable to test new, unseen systems.
In addition, to the best of our knowledge, there is no existing work that aims at automating the conversion of MRs into an executable form, hereafter referred to as EMRs (Executable MRs), which is an essential step to fully automate the MT process.

\textbf{Contribution.}
In this paper, we propose an approach for automatically deriving EMRs from the SUT's requirements using LLMs.
Differently from existing work, our approach does not merely involve querying an LLM for MRs based solely on the LLM knowledge base.
Instead, we rely on prompt engineering practices to teach the LLM both (1)~the specifications of the SUT, which are necessary to derive MRs, and (2)~the DSL to use for EMRs, which is necessary to enable MR execution.

We evaluated the feasibility of our approach through two experiments: (1)~by conducting a questionnaire-based survey in collaboration with Siemens Industry Software, a worldwide leader in providing software and services across industry domains (hereafter, \SISW),
involving four of their software applications, and (2)~by assessing the correctness of the generated EMRs for a Web application.

In our experiments, we used OpenAI GPT-3.5 and GPT-4~\cite{gpt4} as LLM, accessed (to provide prompts and generate responses for deriving MRs and subsequently converting them into EMRs) through the web interface of ChatGPT~\cite{ChatGPT}. 
When converting MRs into EMRs, we instructed the LLM to use SMRL~\cite{Phu:2020} as a domain-specific language (DSL) for specifying MRs in an executable form.

The results are promising, indicating that the generated MRs and EMRs are understandable and relevant for testing, and that the EMRs were correctly converted from their corresponding MRs.

\textbf{Organization.}
The remainder of this paper is structured as follows:
Section~\ref{sec:approach} describes our approach.
Sections~\ref{sec:design} and \ref{sec:results}, respectively, present our experiment design and results.
Sections~\ref{sec:discussion} discusses threats to the validity and research opportunities.
Section~\ref{sec:related work} surveys related work.
Section~\ref{sec:conclusion} concludes the paper. \section{Approach}
\label{sec:approach}

This section describes our approach to automatically deriving EMRs from the requirements of the SUT.
Specifically, our approach relies on an LLM to automate the MT process, particularly by leveraging the LLM's capabilities in understanding and processing natural language documents, identifying relevant information, and synthesizing that information into structured forms (i.e., MRs and EMRs) useful for software testing.
 
\subsection{Overview}
\label{subsec:overview}

\begin{figure}[t]
	\centering	\includegraphics[width=0.8\columnwidth]{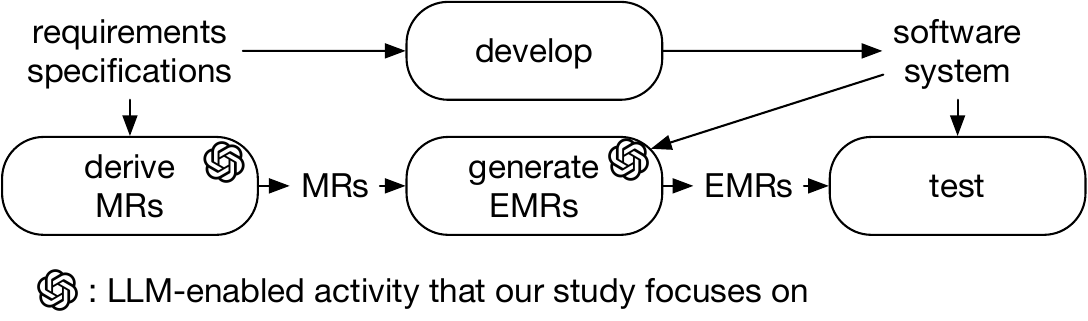}
	\caption{Our approach for generating EMRs using LLMs.}
	\label{fig:overview}
\end{figure}

Our approach, shown in \figurename~\ref{fig:overview}, advances the MT process by introducing two automated activities: ``derive MRs'' and ``generate EMRs''.
The ``derive MRs'' activity takes as input requirements specifications (in  the form of specification documents) and outputs (MRs in natural language).
Unlike existing work that queries an LLM for MRs based on the LLM built-in knowledge, we provide SUT specification documents to the LLM.
This enables our approach to derive SUT-specific MRs, which would otherwise be unattainable because the LLM lacks awareness of the specific software properties that need to be considered.
The ``generate EMRs'' activity takes as input MRs and the SUT (depicted as ``software system'' in Fig.~\ref{fig:overview}).
It outputs EMRs, which are executable programs used to verify that the SUT satisfies the properties captured by the MR.
In the following subsections, we further describe these activities, focusing on their implementation as adopted in our preliminary experiments (see section~\ref{sec:design}).

\subsection{Deriving metamorphic relations}
\label{subsec:deriving MRs}

We aim at automatically deriving MRs from the provided requirements specification documents.
To this end, our approach relies on ChatGPT and develops a prompt sequence for automation.
Since deriving MRs from requirements is a complex task, when using ChatGPT for this purpose, we break the task down into simpler, decomposed steps\footnote{We note that when we passed a single, monolithic query to ChatGPT for deriving MRs, it was unable to derive meaningful MRs.}.

\begin{figure}[t]
	\centering
	\includegraphics[width=\columnwidth]{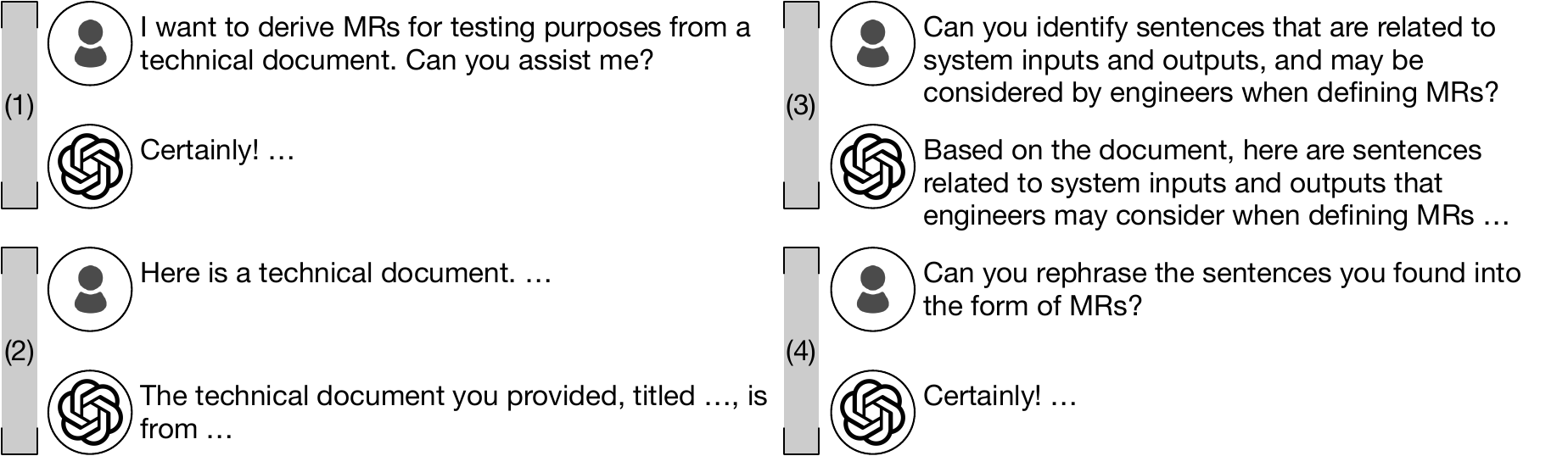}
	\caption{Prompts to derive MRs from the SUT's requirements, grouped by phases in the conversation: (1)~setting the context, (2)~providing requirements specification document(s), (3)~identifying relevant sentences, and (4)~rewriting these sentences into MRs.}
	\label{fig:deriving MRs}
\end{figure}

Figure~\ref{fig:deriving MRs} shows the prompts that derive MRs from the requirements of the SUT.
We have omitted some parts of the text, indicated by ``\dots'', to focus on the essential prompts of our work.
Note that the complete prompts are available online~\cite{REPLICABILITY}.
As shown in Fig.~\ref{fig:deriving MRs}~(1), our prompt sequence begins by setting the context, i.e., deriving MRs, to guide ChatGPT 
for the following interactions.
The remaining sequence, which decomposes the task of deriving MRs, consists of prompts directing ChatGPT to perform simpler steps, as described below.
First, our approach instructs ChatGPT to read a requirements document in order for ChatGPT to access the requirements needed for defining MRs for the SUT (see Fig.~\ref{fig:deriving MRs}~(2)); in response, ChatGPT provides a short summary of it.
Second, given the requirements document, ChatGPT is directed to find sentences that are specifically related to the SUT's inputs and outputs, as well as those that engineers may consider when defining MRs (Fig.~\ref{fig:deriving MRs}~(3)).
This step is important because MRs fundamentally describe how outputs should be changed in response to specific changes in inputs.
The last step then asks ChatGPT to write MRs based on the identified relevant sentences (Fig.~\ref{fig:deriving MRs}~(4)).

For example, consider a scenario where our approach is used to derive MRs from the requirements document of an online shopping system that includes a  \texttt{searchItem} function.
After setting the context and reading the requirements document, ChatGPT could identify requirement R1 in Table~\ref{tbl:req and mr}, which concerns the advanced search options in the \texttt{searchItem} function.
Given requirement R1, ChatGPT then proceeds to define an MR, referred to as MR1 in the table.
MR1 specifies the relationship between the outputs of an initial search query and those of a subsequent query where additional filters are applied. 

\begin{table}[h]
	\caption{An example requirement for an online shopping system and an MR derived from this requirement.}
	\label{tbl:req and mr}
	\centering
	\begin{tabularx}{\columnwidth} {l X}
		\toprule
		R1 & The system should provide advanced search options to allow users to refine their searches based on specific attributes such as price range, category, brand, customer ratings, and availability.\\
		\arrayrulecolor{lightgray}\midrule\arrayrulecolor{black}
		MR1 & For a given search query, applying additional filters (e.g., narrowing down by category or price range) should reduce the number of search results or refine them to match the filters more closely.\\		
		\bottomrule
	\end{tabularx}
\end{table}

\subsection{Generating executable metamorphic relations}
\label{subsec:generating EMRs}

We ask ChatGPT to convert MRs written in natural language into EMRs specified using SMRL~\cite{Phu:2020}, a DSL developed for specifying EMRs.
SMRL provides MR-specific language constructs, which are built on Java, to specify EMRs.
Table~\ref{tbl:SMRL construncts} presents a subset of SMRL constructs.
The construct \texttt{Input(int i)} refers to the i-th sequence of actions performed by a user with the SUT.
The construct \texttt{Output(int i)} refers to the sequence of outputs produced by the SUT in response to \texttt{Input(i)}. 
The construct \texttt{CREATE(Object y, Object x)} defines \texttt{y} as a copy of \texttt{x} and satisfies \texttt{y = x}.
The last three rows in the table capture Boolean expressions corresponding to implication (\texttt{IMPLIES()}), negation (\texttt{NOT()}), and disjunction (\texttt{OR()}).
Using these constructs, engineers can specify EMRs in terms of inputs, input transformations, outputs, output changes, and logical expressions to define the expected relations among them.
In addition, SMRL is already integrated into the MST-wi framework~\cite{Chaleshtari:MST}, which generates executable Java code from specified EMRs to automatically perform MT.

\begin{table}[b]
	\caption{A subset of the SMRL constructs.}
	\label{tbl:SMRL construncts}
	\centering
	\begin{tabularx}{\columnwidth} {@{}l@{}X@{}}
	\toprule
	Construct & Description\\
	\midrule
	\texttt{Input(int i)} & returns the i-th input sequence \\
	\texttt{Output(int i)} & returns the sequence of outputs generate by Input(i)\\
	\texttt{CREATE(Object y,Object x)} & creates y as a copy of x\\
	\texttt{IMPLIES(boolean x,boolean y)} & is equivalent to the Java expression  \texttt{!x||y}\\
	\texttt{NOT(boolean x)} & is equivalent to the Java expression  \texttt{!x}\\
	\texttt{OR(boolean x,boolean y)} & is equivalent to the Java expression  \hbox{\texttt{x||y}}\\	
	\bottomrule
	\end{tabularx}
\end{table}

\begin{figure}[t]
	\centering
	\includegraphics[width=\columnwidth]{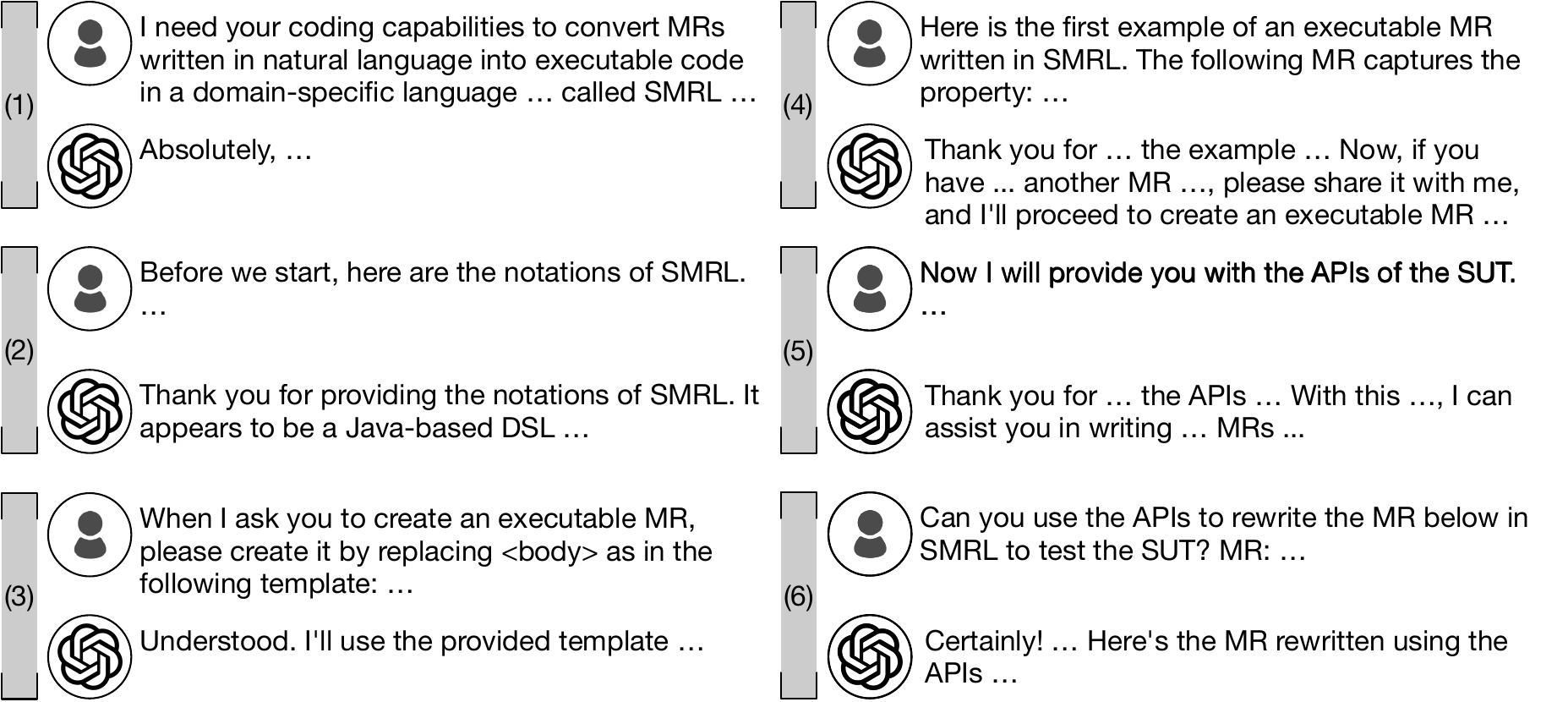}
	\caption{Prompts to convert MRs into EMRs, grouped by phases in the conversation: (1)~setting the context, (2)~learning the syntax of SMRL, (3)~instructing the output format, (4)~few-shot learning of SMRL, (5)~learning the SUT's APIs, and (6)~transforming MRs into EMRs.}
	\label{fig:generating EMRs}
\end{figure}

Figure~\ref{fig:generating EMRs} depicts a sequence of prompts that converts MRs into EMRs, selectively omitting text using ``\dots'' for a clearer exposition.
Our approach involves instructing ChatGPT to understand SMRL and to use SMRL for converting MRs into EMRs.
Specifically, as shown in Fig.~\ref{fig:generating EMRs}~(1), when using ChatGPT, we first set the context by indicating that the subsequent interactions will aim at converting MRs into EMRs.
We then provide ChatGPT with the notations for the SMRL constructs (see Table~\ref{tbl:SMRL construncts}), enabling it to use the SMRL constructs when converting MRs into EMRs (Fig.~\ref{fig:generating EMRs}~(2)).
To transform MRs into EMRs in a consistent form, we provide ChatGPT with an EMR template (Fig.~\ref{fig:generating EMRs}~(3)).
Subsequently, we apply the few-shot learning strategy~\cite{FewShotSurvey} to instruct ChatGPT to understand SMRL, by providing three pairs of MR and corresponding EMR from prior work~\cite{Chaleshtari:MST} (Fig.~\ref{fig:generating EMRs}~(4)).
The few-shot learning strategy is suitable in our study context, as SMRL is a state-of-the-art DSL with a limited number of examples available, and it is built on Java, which ChatGPT already knows.
In addition, we further instruct ChatGPT to understand the SUT's APIs by providing the documentation and signature of the API methods (Fig.~\ref{fig:generating EMRs}~(5)).
For example, to test the Web application described in prior work~\cite{Chaleshtari:MST}, which includes 38 APIs, we need to provide ChatGPT with the details of these APIs.   
Last, we then ask ChatGPT to transform MRs into EMRs using SMRL and to invoke the APIs in the EMRs for testing the SUT (Fig.~\ref{fig:generating EMRs}~(6)).

\begin{figure}
\begin{lstlisting}[style=MR]
MR {{
for (Action searchAction : Input(1).actions()) { //(1)
	if (!isSearchAction(searchAction)) continue; //(2) 
	var originalResults = Output(Input(1), searchAction.getPosition()); //(3)
	for (var filterType : getFilterTypes()) { //(4)
		var filteredInput = applyFilter(Input(1),searchAction.getPosition(),filterType);//(5)
		IMPLIES(
		CREATE(Input(2), filteredInput) && //(6)
		notSameFilterApplied(searchAction, filterType), //(7)
		OR(
		fewerResults(Output(Input(2), searchAction.getPosition()), originalResults), //(8)
		moreRelevantResults(Output(Input(2), searchAction.getPosition()), originalResults, filterType) //(9)
		)
		);//end-IMPLIES
	}//end-for
}//end-for
}}//end-MR
\end{lstlisting}
\caption{An EMR generated by ChatGPT based on the MR presented in Table~\ref{tbl:req and mr}.}
\label{fig:example EMR}
\end{figure}

Figure~\ref{fig:example EMR} shows an EMR generated by our approach, corresponding to the MR described in Table~\ref{tbl:req and mr}.
This EMR specifies that adding filters to a search query should either reduce the number of results or refine them to better match the filter criteria.
Below, we provide the comments generated by ChatGPT to explain the EMR statements; each corresponds to one annotation \texttt{\color{dkgreen}//(\#)} in Fig.~\ref{fig:example EMR}.
\begin{mdframed}
\textit{(1) \& (2): Iterate over all actions within the input sequence to identify search actions.
(3): Retrieve the original search results before any additional filters are applied.
(4): Iterate over possible filter types (e.g., categories, price ranges) that can be applied to the search.
(5): Create a new input sequence where a specific filter is applied to the identified search action.
(6): Ensure the new input sequence (\texttt{Input(2)}) is properly created with the applied filter.
(7): Check that the same filter hasn't already been applied to the search action to avoid redundant checks.
(8): Verify that applying the filter results in fewer search results compared to the original, indicating a refinement of the search output.
(9): Alternatively, check that the results are more relevant to the applied filter, even if the number of results does not decrease. This could involve comparing the characteristics of the search results against the filter criteria to ensure they match more closely.}
\end{mdframed}

We note that we did not provide ChatGPT with the API details of the online shopping system.
Hence, ChatGPT autonomously created some function invocations, such as \texttt{applyFilter()} and \texttt{notSameFilterApplied()}, to implement the EMR using SMRL.
This capability of ChatGPT is desirable in test-driven development practices as it helps engineers identify which APIs are needed to test the SUT at early development stages. \section{Experiment Design}
\label{sec:design}

To assess the feasibility of our approach, we have carried out two experiments, namely EXP1 and EXP2.
EXP1 aims at collecting feedback from practitioners.
EXP2 aims at evaluating the correctness of generated EMRs.

\textbf{EXP1 - \SISW{} applications.}
To assess whether our approach can practically help software practitioners, it is important to collect and analyze their perceptions on our work.
In collaboration with \SISW, we have conducted a questionnaire-based survey to gather their feedback on the MRs and EMRs obtained by applying our approach to their software applications.
Due to confidentiality reasons, \SISW{} was not able to share the requirements and the APIs of their applications; instead, \SISW{} selected four non-confidential technical documents based on their interests. These documents describe their applications (modelling and simulation) for marine design~\cite{SIEMENSmarine}, wind turbine~\cite{SIEMENSwind}, aircraft propulsion~\cite{SIEMENSpropulsion}, and aero-acoustics~\cite{beriot2015high}.
They provided us with these documents (83 pages in total) as requirements specifications.
As remarked above, in absence of appropriate APIs, ChatGPT suggests invoking functions that should be implemented to automate the execution of the MR.
Further, despite the lack of APIs, considering the preliminary nature of this work, these subjects still enable us to gather feedback from practitioners on how well ChatGPT can generate EMRs.

In EXP1, we relied on GPT-4 because, at the time we designed our experiments (January 2024), it was the only 
available top-5 leaderboard LLM capable of processing PDF documents\footnote{\url{https://lmsys.org/blog/2023-12-07-leaderboard/}}.
Note that our questionnaire-based survey study does not enforce practitioners to respond to all questions.
We adopted this survey strategy to obtain higher quality responses to the questions that practitioners feel confident in answering, since the participating practitioners have varying levels of expertise across the four applications developed by \SISW.

The questionnaire for MRs (resp. EMRs) contains three statements for each MR (resp. EMR).
It requests practitioners to indicate their level of agreement with the statements on a Likert scale (i.e., strongly agree, agree, neutral, disagree, and strongly disagree).
We designed the statements in our questionnaire-based survey, drawing inspiration from Rogers’ theory of innovation diffusion~\cite{rogers2014diffusion}.
In this theory, the following five characteristics, based on practitioners' perceptions, are introduced for their impact on the adoption of innovative solutions:
\begin{enumerate}
	\item [–] Complexity: Reflecting the extent to which an innovation is perceived as challenging to understand or implement.
	\item [–] Compatibility: Referring to the extent to which an innovation aligns with practitioners' existing values, experiences, and needs.
	\item [–] Trialability: Indicating the extent to which an innovation can be tried on a limited scale or adopted incrementally.
	\item [–] Observability: Pertaining to the visibility of the results of an innovation to others.
	\item [–] Relative advantage: Describing the perception of an innovation's superiority compared to what is currently used.
\end{enumerate}
Out of these five characteristics, our focus is on the first three: complexity, compatibility, and trialability.
Note that the last two characteristics remain unaddressed in our survey, as our approach is still preliminary and has not yet been deployed in practice.

The questionnaire on MRs, available online~\cite{REPLICABILITY}, includes, for each MR, a form showing the generated MR (e.g., see Table~\ref{tbl:req and mr}) and three assessment statements (S1, S2, and S3) described in Table~\ref{tbl:MR statements}.
These forms are grouped by application, and for each application, the questionnaire invites practitioners to provide open feedback.
Statement S1 in Table~\ref{tbl:MR statements} is about complexity; it assesses the degree to which a derived MR is understandable.
Statement S2 relates to compatibility; it assesses the degree to which a derived MR aligns with practitioners' perceptions as a property to be considered in testing the SUT.
Statement S3 concerns triability; it examines how helpful a derived MR is as an instrument for defining test cases (i.e., input and expected output pairs).

\begin{table}[t]
	\caption{Statements (S1, S2, and S3) for collecting feedback on MRs.}
	\label{tbl:MR statements}
	\centering
	\footnotesize
	\begin{tabularx}{\columnwidth} {l X}
		\toprule
		S1 & The MR is easy to understand. \\
		& $\square$ strongly agree  \quad  $\square$ agree \quad $\square$ neutral \quad $\square$ disagree \quad $\square$ strongly disagree \\
		\arrayrulecolor{lightgray}\midrule\arrayrulecolor{black}
		S2 & The MR captures a property that needs to be considered when testing the application. \\
		& $\square$ strongly agree  \quad  $\square$ agree \quad $\square$ neutral \quad $\square$ disagree \quad $\square$ strongly disagree \\
		\arrayrulecolor{lightgray}\midrule\arrayrulecolor{black}
		S3 & The MR helps identify the expected outputs for given inputs. \\
		& $\square$ strongly agree  \quad  $\square$ agree \quad $\square$ neutral \quad $\square$ disagree \quad $\square$ strongly disagree \\
		\bottomrule
	\end{tabularx}
\end{table}

The questionnaire on EMRs, available online~\cite{REPLICABILITY}, first introduces SMRL by explaining its constructs, providing three examples of EMRs written in SMRL, and a link to the SMRL paper~\cite{Chaleshtari:MST}.
The questionnaire then, for each MR, provides a form with an MR (e.g., see Table~\ref{tbl:req and mr}), an EMR (e.g., see Fig.~\ref{fig:example EMR}), and three statements (S1$^E$, S2$^E$, and S3$^E$, listed in Table~\ref{tbl:EMR statements}).
These forms are grouped by each \SISW{} application.
For each application group, the questionnaire includes an open question to collect additional feedback.
Statement S1$^E$ in Table~\ref{tbl:EMR statements} is about complexity, assessing the extent to which an EMR is understandable.
Statement S2$^E$ relates to compatibility, evaluating the degree to which an EMR aligns with its corresponding MR.
Statement S3$^E$ concerns triability, examining the feasibility of implementing the functions invoked in an EMR.

\begin{table}[b]
	\caption{Statements (S1$^E$, S2$^E$, and S3$^E$) for collecting feedback on EMRs.}
	\label{tbl:EMR statements}
	\centering
	\footnotesize
	\begin{tabularx}{\columnwidth} {l X}
		\toprule
		S1$^E$ & The EMR is easy to understand. \\
		& $\square$ strongly agree  \quad  $\square$ agree \quad $\square$ neutral \quad $\square$ disagree \quad $\square$ strongly disagree \\
		\arrayrulecolor{lightgray}\midrule\arrayrulecolor{black}
		S2$^E$ & The EMR is consistent with the corresponding MR. \\
		& $\square$ strongly agree  \quad  $\square$ agree \quad $\square$ neutral \quad $\square$ disagree \quad $\square$ strongly disagree \\
		\arrayrulecolor{lightgray}\midrule\arrayrulecolor{black}
		S3$^E$ & It is feasible to implement all the functions used in the EMR. \\
		& $\square$ strongly agree  \quad  $\square$ agree \quad $\square$ neutral \quad $\square$ disagree \quad $\square$ strongly disagree \\
		\bottomrule
	\end{tabularx}
\end{table}

Our approach derived 50 MRs and their corresponding EMRs from the four documents.
Of these MRs, \SISW{} independently assigned two practitioners to evaluate the 14 MRs derived for the aero-acoustics application, while one was assigned to 36 MRs for the remaining three applications, totaling 64 MRs analyzed.
For the EMRs, one practitioner was assigned to all the applications. 
Note that these assignments were independently decided by \SISW.

\textbf{EXP2 - Web application.}
EXP2 evaluates the correctness of the EMRs generated by our approach.
Recall that EXP1 studies EMRs that are incomplete with regard to invoking the SUTs' APIs.
Therefore, EXP2 employs a different application, for which we have its API specifications.
EXP2 uses the experiment dataset provided by the prior work on MST-wi~\cite{Chaleshtari:MST}, containing both the APIs enabling the testing of a Web application (hereafter, SUT) and MRs that are designed to detect security vulnerabilities.
Specifically, in EXP2, we randomly selected ten MRs among the ones published online \emph{after the cutoff date} of GPT-3.5 (in EXP2, we relied on GPT-3.5 for this reason).
Given the ten MRs and the SUT's APIs, we applied our approach to produce ten corresponding EMRs.

\begin{table}[t]
	\caption{Annotation labels used in EXP2.}
	\label{tbl:annotation labels}
	\centering
	\begin{tabularx}{\columnwidth} {l l X}
		\toprule
		Label  & Type & Description \\
		\midrule
        CLC & Simple & The statement contains a correct language construct. \\
        \arrayrulecolor{lightgray}\midrule\arrayrulecolor{black}
		C & Complex & The statement is correct. \\
		\arrayrulecolor{lightgray}\midrule\arrayrulecolor{black}
		AI & Complex & The statement does not perform exactly what the original MR does, but it is a valid alternative implementation. \\
        \midrule
        WLC & Simple & The statement misuses a language construct.\\
		\arrayrulecolor{lightgray}\midrule\arrayrulecolor{black}
		WS & Complex & The statement is wrong; it misses some required actions, performs the wrong operation, and the explanation does not reflect what is requested in the MR text. \\
		\arrayrulecolor{lightgray}\midrule\arrayrulecolor{black}
		WI & Complex & The statement does not correctly implement what is suggested in the explanation. \\
		\arrayrulecolor{lightgray}\midrule\arrayrulecolor{black}
		IE & Complex & The statement invokes an API function invented by ChatGPT, even if an adequate one exists. \\
		\arrayrulecolor{lightgray}\midrule\arrayrulecolor{black}
		INE & Complex & The statement invokes an API function invented by ChatGPT because an adequate one does not exist. \\
		\arrayrulecolor{lightgray}\midrule\arrayrulecolor{black}
		ITE & Complex & The statement invokes an API function invented by ChatGPT but delegates too much logic to it. \\
		\arrayrulecolor{lightgray}\midrule\arrayrulecolor{black}
		ES & Complex & The statement invokes an existing and appropriate API function but swaps parameters. \\
		\arrayrulecolor{lightgray}\midrule\arrayrulecolor{black}
		ENO & Complex & The statement invokes an existing and appropriate API function but not in an object-oriented manner. \\ 
		\arrayrulecolor{lightgray}\midrule\arrayrulecolor{black}
		WAU & Complex & The statement misuses valid APIs. \\
		\arrayrulecolor{lightgray}\midrule\arrayrulecolor{black}
		MISS & Complex &The statement misses an instruction required to implement what is reported in the explanation.\\ \bottomrule
	\end{tabularx}
\end{table}

To assess the correctness of the ten generated EMRs, we inspected and annotated each line of the EMRs with one of the 13 labels described in Table~\ref{tbl:annotation labels}.
In the annotation process, we considered both the EMRs and their explanations (i.e., comments on the EMRs) generated by our approach.
Note that each EMR statement is associated with an explanation in natural language produced by ChatGPT.
Our labels distinguish between \emph{complex} and \emph{simple} statements. Simple statements are the ones including only one language construct (e.g., \texttt{IMPLIES}, \texttt{NOT}, \texttt{OR}). Complex statements include at least one method invocation. 
Further, our labels  are classified into two groups: \emph{correct} and \emph{incorrect}.
The former contains three labels (i.e., C, CLC, and AI in Table~\ref{tbl:annotation labels}) indicating, respectively, whether a statement in an EMR  implements part of the corresponding MR exactly, uses SMRL constructs correctly, or implements valid alternatives.
The latter includes the remaining ten labels, each indicating a different type of issue, such as ``incorrect use of SMRL constructs'' and ``misuse of the SUT's APIs''.
For example, in Fig.~\ref{fig:example EMR}, the statement \texttt{if (!isSearchAction(searchAction)) continue;} on line 3 is annotated with the INE label, indicating that ChatGPT invented \texttt{isSearchAction()} and the EMR invokes it because an adequate API of the SUT was not provided to ChatGPT when generating the EMR from the MR described in Table~\ref{tbl:req and mr}. 

\section{Experiment results}
\label{sec:results}

\textbf{EXP1.}
Table~\ref{tbl:EXP1 results} shows the results of EXP1, focusing on the combined number of responses (i.e., Likert ratings) for each statement across the 64 MRs reviewed by the three practitioners at \SISW.
From the results of statement S1, we found that in 49 out of the 64 MRs (77\%), the practitioners were able to understand the proposed MRs and expressed positive feedback, i.e., ``strongly agree'' or ``agree''.
In 12 MRs, the practitioners were neutral, and in the remaining 3 MRs, they disagreed with statement S1.
Regarding statement S2, the practitioners agreed  that the MRs capture properties that need to be considered when testing the SUT for 41 out of the 64 MRs (64\%).
In 10 MRs, the practitioners were neutral; in the remaining 13 MRs, the practitioners disagreed or strongly disagreed with statement S2.
In contrast to these results, which are more positive, the responses to statement S3 showed mixed ratings.
This statement indicates whether the presented MR helps identify the expected outputs for given inputs.
In 18 MRs, the practitioners agreed; in 19 MRs, they were neutral; and in the remaining 27, they disagreed or strongly disagreed.

\begin{table}[b]
	\caption{Responses to the MRs survey, for each statement (S1, S2, S3) in Table~\ref{tbl:MR statements}.}
	\label{tbl:EXP1 results}
	\centering
	\footnotesize
\begin{tabularx}{\columnwidth} {@{}l Y@{\hskip-1em}Y@{\hskip-1em}Y@{\hskip-1em}Y@{\hskip-1em}Y@{}}
		\toprule
		& strongly agree & agree & neutral & disagree & strongly disagree \\
		\midrule
		S1 & 3 & 46 & 12 & 3 & 0 \\
		S2 & 0 & 41 & 10 & 12 & 1 \\
		S3 & 0 & 18 & 19 & 24 & 3 \\		
		\bottomrule
	\end{tabularx}\end{table}

Regarding the non-positive responses to statement S3, we identified possible reasons from the practitioners' qualitative feedback.
In particular, one practitioner, who provided 11 ratings as ``disagree'' and one as ``strongly disagree'' to the statement, stated that while most of the MRs are correct and useful for reminding someone what to test, they are too generic to aid in deriving appropriate testing.
This aligns with feedback from another practitioner at \SISW{} who, despite not assessing the MRs, collected and reviewed the responses to the MRs questionnaire.
The practitioner provided a possible reason for some of the MRs being very generic, stating that it could be attributed to the nature of the documents provided by \SISW.
 
Regarding the survey results for the 50 EMRs in EXP1, one practitioner provided consistent responses across all EMRs, stating they
perceived that the EMRs \emph{were clean and understandable}
and \emph{consistent with their corresponding MRs}.
However, the practitioner deemed that not all functions used in the EMRs were feasible to be implemented.
Regarding the last opinion, the practitioner noted that the functions in the EMRs, generated by ChatGPT, were specific to each EMR.
Therefore, there is little opportunity for reusing these functions across different EMRs, meaning the effort required for implementation would be significant.
In addition, the specific constraints of the testing environment for their applications may pose challenges in using the exposed APIs to implement functions in the EMRs.
Even though EXP1 generated incomplete EMRs with regard to using the SUT's APIs, as discussed above, this finding suggests that the characteristics of the SUT, such as its APIs, should be considered when further elaborating our approach to deriving EMRs.

\begin{tcolorbox}[left=2pt,right=2pt,top=0pt,bottom=0pt]
	The results for EXP1 indicate that our approach generates MRs and EMRs that practitioners deem to be understandable and relevant for testing.
\end{tcolorbox}

\textbf{EXP2.}
In EXP2, we annotated each line of the ten EMRs, which were converted from the corresponding ten MRs studied in prior work~\cite{Chaleshtari:MST}.
These EMRs have a minimum of 10 statements, a mean of 13.6 statements, and a maximum of 20 statements, totaling 136 statements.

 \begin{table}[b]
 	\caption{Distribution of annotation labels (from Table~\ref{tbl:annotation labels}) for the 10 EMRs obtained from EXP2.}
 	\label{tbl:EXP2 results}
 	\centering
 	\footnotesize
 	\begin{tabularx}{\columnwidth} {Y Y Y Y Y Y Y Y Y Y Y Y Y Y Y}
 		\toprule
 		C & CLC & AI & WS & WI & IE & INE & ITE & ES & ENO & WAU & WLC & MISS \\
 		\midrule
 		52 & 54 & 1 & 3 & 0 & 3 & 1 & 9 & 1 & 2 & 0 & 10 & 3 \\
        38.2\%&   39.7\%& 0.7\% &2.2\% &0.0\% &0.7\% &0.0\% &6.6\% &0.7\% &1.5\% & 0.0\% &7.4\% &2.2\%\\
 		\bottomrule
 	\end{tabularx}
 \end{table}

Table~\ref{tbl:EXP2 results} shows the results of EXP2 annotations using the labels defined in Table~\ref{tbl:annotation labels}.
Out of the 136 statements in the EMR code, the majority (107 statements, 78.6\%) were classified as correctly converted.
Specifically, 52 of the 107 statements are complex, and correctly and exactly implement part of the corresponding MRs, as shown by column C in Table~\ref{tbl:EXP2 results}.
54 of the 107 statements are simple, and correctly use SMRL constructs (column CLC).
The remaining complex statement is valid and correct with respect to the corresponding MR (column AI). 
As for the 30 statements that were deemed incorrect, they can be further categorized as follows:
\begin{itemize}

\item 16 statements incorrectly use APIs (labeled with IE, INE, ITE, ES, and ENO). For example, five ITE cases introduce the method \texttt{isAuthorized} to check if a user is authorized to perform an action, instead of verifying that the outputs obtained by the source and the follow-up inputs differ.
\item 10 statements incorrectly use SMRL constructs (column WLC). They are all related to the LLM ``forgetting'' how to use the construct \texttt{IMPLIES}; indeed, instead of separating the right-hand side of the implication with a comma (see end of Line 9 in Fig.~\ref{fig:example EMR}), it uses an \texttt{\&}. They can be easily fixed manually.
\item Three statements are wrong (column WS); for example, the generated MR checks if the returned page is an error page instead of the opposite. 
\item Three statements miss implementing what is described in the corresponding explanations (column MISS); specifically, they do not compare the outputs of the source and the follow-up inputs.
 Further, these three statements had been annotated also with another label because, in addition to miss some instructions, they contained an incorrect implementation (i.e., ENO and WS), which is the reason why our labels sum up to 139 instead of 136.
\end{itemize}

\begin{tcolorbox}[left=2pt,right=2pt,top=0pt,bottom=0pt]
 	The results for EXP2 indicate that the generated EMRs are largely correct with respect to their MRs.
 \end{tcolorbox}
 
 \section{Threats to Validity and Research Opportunities}
 \label{sec:discussion}
 
Even though the experiment results are promising, we recognize that our work is in its preliminary stages, and there are some potential threats to its validity.
In this section, we discuss these concerns, focusing on those that lead to challenging yet important directions for future research.
 
\textbf{Prompt engineering.}
Our prompts are designed by decomposing the processes of deriving MRs from requirements and transforming MRs into EMRs into smaller, simpler steps.
Despite these decomposed steps aligning with those that engineers would conduct manually, when using LLMs there might be better sequences of prompts that could improve the accuracy and relevance of the automatically generated MRs and EMRs.
Further research is necessary to refine and compare the prompts, ensuring that they efficiently guide LLMs in producing effective MRs and EMRs.

\textbf{Assessing MRs and EMRs.}
We evaluated the MRs and EMRs generated by our approach using the questionnaire-base survey and annotation studies, which rely on the quality of human assessments.
Hence, such assessments are naturally biased by the participating personnel in the study.
To minimize any threat to construct validity (i.e., researcher expectations affecting human subjects' assessment), our industrial partner, \SISW, invited four practitioners with different backgrounds for the questionnaire-based survey study.
For the annotation study, the second author, who is familiar with SMRL, annotated the EMRs to ensure that the annotations are accurate. Although our choice may introduce a construct validity threat, the availability of experimental data enables other researchers to further investigate our findings.

As part of our future work, to better address face validity (i.e., the selection of appropriate reflective indicators), we aim to leverage recent studies on the assessment of code generation models~\cite{MetricsStudy} and select additional quantitative metrics to measure the degree to which MRs and EMRs are accurate and relevant to the SUT.
Such metrics are desirable not only for objective assessment but also for automating the MT process; indeed, such metrics might be used as reward to improve LLMs results.

\textbf{Human expertise.}
In our work, 
we leveraged human expertise
to select technical documents and assess the outputs, i.e., MRs and EMRs.
Since our approach does not involve humans-in-the-loop (e.g., the textual MRs generated by the LLM for EXP1 are not rewritten by humans before generating EMRs), 
the quality of outputs highly depends on the quality of inputs.
To mitigate any bias originated from the input documents (construct validity), \SISW{} selected four technical documents independently from the authors who developed the approach.

In the future, 
even if we aim to automate the MT process, incorporating humans in the loop could be use used to refine and validate the inputs and the intermediate outputs, to efficiently and effectively guide our approach to produce more accurate and relevant MRs and EMRs.
Therefore, future research should explore methods to effectively incorporate human expertise into an AI-based MT process, balancing automation with human interventions.

\textbf{External validity.} To mitigate generalizability threats, for EXP1, we considered specifications of industrial modelling and simulation software for four different applications domains (marine design, wind turbine, air-craft propulsion, and aero-acoustics). Although such software may differ from other types of software systems (e.g., content management software), it is of high complexity and adopted in many industrial sectors including automotive and space. 
Further, for EXP2, we focused on MRs that demonstrated effective for content management (e.g., Joomla~\cite{Joomla}) and Web-based automation software (e.g., Jenkins~\cite{Jenkins}), thus complementing the choice made for EXP1. \section{Related Work}
\label{sec:related work}

Many MT approaches have been developed for testing various SUTs, such as search engines~\cite{Zhou2016}, Web applications~\cite{Chaleshtari:MST}, and embedded systems~\cite{Ayerdi:2020}, accounting for the requirements specific to these SUTs.
However, these approaches heavily rely on manual efforts to elicit MRs, inherently limiting the applicability of MT.
To reduce the cost of defining MRs, researchers have proposed relying on meta-heuristic search~\cite{ZCH+14,ZLDS+19} and genetic programming~\cite{Terragni} to automatically derive MRs from execution data.
Although promising, such approaches require several executions of the SUT, which makes them suitable for small programs, but may not scale for large software where executions may take several minutes and a lot of outputs are produced.

Recently, a few studies~\cite{luu2023chatgpt,Zhang:MRs:2023} have proposed the use of LLMs to automatically derive MRs from the knowledge base of LLMs.
However, these studies are limited to deriving MRs for SUTs already known in the adopted LLMs.
In other words, there are  no studies demonstrating the applicability of LLMs to derive MRs that account for the requirements specific to an SUT that was unseen during the LLM training phases. Considering that, in industrial contexts, software testing activities often target new products implementing requirements different from those implemented by existing systems, LLM-based approaches for the automated generation of MRs should be capable of handling unseen requirements.
This capability is what we presented in this paper.

Additionally, we note that, to the best of our knowledge, there is no work that automates the conversion of MRs into an executable form, which is required to fully automate the MT process.
Chaleshtari et al.~\cite{Chaleshtari:MST} proposed the adoption of a DSL to automate the execution of MRs; however, DSL-based MRs may have limited readability, contrary to MRs in natural language.

In summary, there is no solution in the literature that aims at fully automating the MT process, involving the derivation of MRs from SUT-specific requirements and the generation of EMRs, for the automated determination of test results in MT.
An alternative is the automated generation of test cases---including oracles---using LLMs, which is under active development~\cite{wang2024software}.
However, since LLM-generated code may still require human validation, we believe that validating EMRs is more cost-effective than validating test cases. 
Indeed, a single validated EMR enables exercising the SUT with multiple inputs, which might otherwise be exercised by different test cases, each needing validation.
 \section{Conclusion and Future work}
\label{sec:conclusion}

In this paper, we have introduced our approach for automatically deriving EMRs from requirements using LLMs.
In our preliminary experiments, we applied our approach to four applications of our industry partner (\SISW{}) and conducted a questionnaire-based survey study to collect opinions from \SISW{} practitioners.
In addition, we evaluated the correctness of the EMRs generated by our approach, by applying it to a Web application.
The feedback from practitioners at \SISW{} confirmed that our work is a promising direction for automating the derivation of MRs and EMRs from requirements.
Furthermore, our annotation results indicate that our approach has strong potential for correctly generating EMRs from MRs.
However, our results  also reveal that the derived MRs need to be more specific to the SUTs, and the derived EMRs could be further improved with regard to the use of DSL constructs and APIs, which necessitate further research.  

In our future work, we plan to explore several key areas related to our current research.
A primary focus will be on automating prompt engineering to facilitate the generation of prompts for deriving MRs from requirements and their subsequent conversion into EMRs.
Recognizing the significance of input quality, particularly in requirements, another important area will involve developing methods to assist engineers in providing high-quality inputs.
Integrating human expertise, or ``human in the loop'', will also be an important aspect of our research, aiming to further enhance the accuracy and relevance of the generated MRs and EMRs.

\textbf{Data availability.}
Our experiment package with prompts, MRs, EMRs, and questionnaires is available online~\cite{REPLICABILITY}.

\bibliographystyle{plain}

\end{document}